\documentstyle[preprint,aps]{revtex}

\begin{document}

\preprint{\vbox{\hbox{ }}}
\title{Exact Scattering States of Dirac-Born-Infeld Equation with 
       Constant Background Fields}
\author{Chuan-Tsung Chan}
\address{Department of Physics, National Taiwan University, 
         Taipei, Taiwan, 106} 
\date{\today}
\maketitle

\begin{abstract}
   
  Exact solutions to the Dirac-Born-Infeld equation, which describes scatterings
of localized wave packets in the presence of constant background fields, are 
derived in this paper.  
                                                                 
\end{abstract}

\vspace{2cm}

\pacs{PACS numbers:}

\newpage

\section{Introduction}                                                         

  The Born-Infeld action was first proposed by Born and Infeld 
\cite{Born:1934gh} as a finite limiting field modification of the classical 
Maxwell theory of electromagnetism.
  It was later generalized by Dirac \cite{Dirac:1962iy} as an extensible 
model of electrons, with a profound assumption that different flavors, e.g., 
muon, correspond to membrane excitations coupled to the Born-Infeld action.
  While successive efforts trying to put through a fully quantized 
version of this classical theory \cite{deWit:1988ig} only receive their revival
after D-brane \cite{Polchinski:1995mt} and matrix model \cite{Banks:1997vh} 
\cite{Susskind:1997cw} revolutions, the dynamical properties of this 
theory were largely unexplored.   
  With the advent of D-brane technology in string theory, the Dirac-Born-Infeld
Lagrangian has been identified as the low energy effective theory describing 
the interactions of massless excitations in the brane world 
\cite{Polchinski:1996na}.
  In view of this progress, many research papers have focused on the symmetry 
properties, e.g., string duality, and qualitative behaviors of D-branes
within special kinematic regions and parameter space.
  Nevertheless, detailed dynamical solutions of this theory remain illusive.
  Given that the solitonic nature of D-branes in string theory is established
semiclassically, it is prudent to study exact solutions of the classical 
theory, namely, the Dirac-Born-Infeld equation, as a first approximation.
  Hopefully, with either supersymmetry constraints and/or small quantum 
corrections, which protect classical analysis from quantum fluctuations, 
we can uncover certain nonperturbative aspects of string dynamics with these 
solutions.
  
  In this paper, I shall focus on a particular class of exact solutions to 
the Dirac-Born-Infeld equation, which generalizes previous work by
Barbashov and Chernikov \cite{Barbashov:1966wb} \cite{Barbashov:1967wb} on 
the wave solutions to the Born-Infeld theory. 
  These solutions describe scatterings among localized wave packets propagating
along D-string in a two-dimensional plane under the influence of constant 
background fields, including both a second rank antisymmetric tensor 
$B_{\mu \nu}$ and a $U(1)$ gauge field strength tensor ${\hat F}_{ij}$.
  In addition to the study of time evolution for these scattering events, 
I also examine the connections between phase shifts and total energy momentum
of the D-string.

  Part of the motivation leads to the present study is tied to the recent
discussion on various connections between Dirac-Born-Infeld theory and fluid 
dynamics \cite{Jackiw:1999wy} \cite{Baker:1999bf}. 
  While the original derivation by Barbashov and Chernikov is based on 
geometrical arguments, I opt for an analytical approach by solving partial 
differential equations.
  In this regard, the nature of wave solutions is studied purely within a 
classical field theoretical context.

  This paper is organized as follows: Basic properties of the Dirac-Born-Infeld
theory, together with several special solutions are reviewed in Sec. II. 
  Then we solve the Dirac-Born-Infeld equation with constant background fields
of the D-string in Sec. III.
  After some discussions about the scattering event and the physical 
interpretation of phase shifts, we give summary and conclusion in Sec. IV. 
     
\section{Basic Properties of the Dirac-Born-Infeld Equation with 
         Constant Background Fields} 

\subsection{General background of the DBI theory}
  
  The Dirac-Born-Infeld (DBI) Lagrangian for a Dp-brane is given by
\begin{equation}
 {\cal L} = \sqrt{ (-1)^p \det [ \hat G + \hat B + \hat F ]_{ij} },
 \hspace{0.7cm} i, j = 0,1,...,p,
\label{eq:dbi0}
\end{equation}
where 
\begin{equation}
 {\hat G}_{ij} \equiv \frac{\partial X^\mu}{\partial x_i} 
                      \frac{\partial X^\nu}{\partial x_j} 
                          G_{\mu\nu}(X), 
  \hspace{1.5cm} \mu, \nu = 0,1,.....,9,
\end{equation}
is the induced metric on the brane world-volume, and 
\begin{equation}
 {\hat B}_{ij} \equiv \frac{\partial X^\mu}{\partial x_i}
                      \frac{\partial X^\nu}{\partial x_j}
                       B_{\mu\nu}(X),
 \hspace{1.5cm} \mu, \nu = 0,1,.....,9,
\end{equation}
describes the interaction of the Dp-brane with the antisymmetric tensor field
$B_{\mu\nu}(X)$, which is generated from the close string spectrum.
  Finally,
\begin{equation} 
 {\hat F}_{ij}(x) \equiv  
   \frac{\partial A_j}{\partial x_i} 
 - \frac{\partial A_i}{\partial x_j},
 \hspace{2.2cm} i, j = 0,1,.....,p, 
\end{equation}
is a $U(1)$ Abelian field strength tensor, which arises from the open string 
modes $A_i$ on the Dp-brane \cite{Dai:1989ua} 
\cite{Leigh:1989jq},
\begin{equation}
  A(x) \equiv A_i (x) \ d x^i.
\end{equation} 

  One special feature associated with the DBI action is that
it is invariant under general coordinate transformations,
\begin{equation}
 x'_i = x'_i (x_0, x_1,....., x_p), \hspace{1cm} i=0,1,.....,p,
\end{equation}
with
\begin{equation}
 d x'_j = \left( \frac{\partial x'_j}{\partial x_k} \right) d x_k
 \equiv J_{jk} \ d x_k,
\hspace{1cm}
          A_i =  A'_j  
    \left( \frac{\partial x'_j} {\partial x_i} \right) \equiv 
          J_{ij}^t A'_j,
\end{equation}
such that
\begin{equation}
 {\cal L}' = \left( \frac{\partial \{ x_0, x_1, ...., x_p \} }
                               {\partial \{ x'_0, x'_1, ...., x'_p \}}
                                \right) {\cal L} \ \Rightarrow \
              S' = \int d^{p+1} x'\ {\cal L}'
                 = \int d^{p+1} x' J^{-1} \ {\cal L}
                 = \int d^{p+1} x \ {\cal L} = S.
\end{equation}
  Consequently, the equation of motion associated with the DBI action is 
reparametrization invariant.

  On the other hand, there are two gauge transformations which leave the DBI
Lagrangian invariant,
\begin{eqnarray}
 &(i)& \hspace{0.3cm} \delta A_i = \partial_i \lambda, \ \
       \delta B_{\mu\nu} = 0 \hspace{3.7cm}
       \Rightarrow \ \hat F' = \hat F', \ \hat B'= \hat B ; 
\label{eq:gau1} \\
 &(ii)& \hspace{0.3cm} \delta A_i = - \zeta_\mu 
       \left( \frac{\partial X^\mu}{\partial x^i} \right), \ \
       \delta B_{\mu\nu} = \partial_\mu \zeta_\nu - \partial_\nu \zeta_\mu \
       \Rightarrow \ \hat B' + \hat F' = \hat B + \hat F.
\label{eq:gau2}
\end{eqnarray}
  In view of these symmetries, not all components of $B_{\mu\nu}$ and ${\hat 
F}_{ij}$ are independent, and physical quantities should depend only on the 
gauge invariant combination $\hat B + \hat F$. 

\subsection{DBI equation for the D-string}

  In this paper, we shall focus on a special case, where a D-1 brane (or 
D-string) is moving within a two-dimensional plane of the nine-dimensional 
space. 
  That is, we choose the world-sheet coordinates of the D-string to be
\begin{equation}
  X_0 = x_0 = t, \hspace{2cm} X_1 = x_1 = x, 
\end{equation}
and the profile of the D-string is described by 
\begin{eqnarray}
  X_2 &=& X_2 (x_1, x_0) \equiv \phi (x, t), \\
  X_3 &=& X_4 = ..... = X_9 = 0.
\end{eqnarray}

  In a flat space-time metric
\begin{equation}
 G_{\mu\nu}(X) = \mbox{diag }(-1,1,1,.....,1),
\end{equation}
an expansion of the determinant inside the DBI Lagrangian, 
Eq.(\ref{eq:dbi0}), gives 
\begin{equation}
{\cal L} = \sqrt{ (1  -  B_3^2) 
                - (1  +  B_1^2) \phi_t^2 
                -  2 B_1 B_2    \phi_t \phi_x 
                + (1  -  B_2^2) \phi_x^2 
                +  2 B_1 B_3    \phi_t
                +  2 B_2 B_3    \phi_x },
\label{eq:lag}
\end{equation}
where we have introduced the external field variables
\begin{equation}
 B_1 \equiv B_{12}, \hspace{1cm}
 B_2 \equiv B_{20}, \hspace{1cm}
 B_3 \equiv B_{01} + F_{01}.
\end{equation}
  We shall assume that $B_i$ are constants
and one can verify that the Lagrangian Eq.(\ref{eq:lag}) 
is invariant under gauge transformations Eqs.(\ref{eq:gau1}) (\ref{eq:gau2}). 

  If we perform a Wick rotation,
\begin{equation}
 t = i \tau, \hspace{1cm} B_1 = i b_1, 
\end{equation}
then the Lagrangian becomes 
\begin{equation}
{\cal L} = \sqrt{ (1 - B_3^2) 
               +  (1 - b_1^2) \phi_\tau^2
               - 2 b_1 B_2    \phi_\tau \phi_x
               +  (1 - B_2^2) \phi_x^2
               + 2 b_1 B_3    \phi_\tau
               + 2 B_2 B_3    \phi_x }.
\end{equation} 
From this form, we observe a discrete symmetry associated with the DBI
Lagrangian of the D-string in the presence of background fields, 
\begin{equation}
  - i t \leftrightarrow x, \hspace{1cm} - i B_1 \leftrightarrow B_2.
\end{equation}  
  
  The Euler-Lagrange equation for the Dirac-Born-Infeld action in the 
presence of constant background fields $B_i$ for the D-string is given by
a second order nonlinear partial differential equation,
\begin{eqnarray}
     [ (1 - B_2^2 -     B_3^2) + 2 B_1     B_3        \phi_t 
   - \ (1 + B_1^2 -     B_2^2)  \phi_t^2 \ \ \ ] \    \phi_{xx} & &  
     \nonumber \\
     + [- 2 B_1 B_2 - 2 B_1 B_3 \phi_x + 2 B_2 B_3    \phi_t 
   + 2 (1 + B_1^2 -     B_2^2)  \phi_x     \phi_t ] \ \phi_{xt} & &  
     \nonumber \\
 + [ - (1 + B_1^2 -     B_3^2) - 2 B_2 B_3 \phi_x 
   - \ (1 + B_1^2 -     B_2^2)  \phi_x^2 \ \ \ ] \    \phi_{tt} &=& 0.
\label{eq:dbi} 
\end{eqnarray}

\subsection{Special cases of the DBI equation with 
            constant background fields}

  Before we solve the partial differential equation for D-string, 
let us make some simple observations:
  If $B_1 = B_2 = 0$ and $B_3^2 <1$, the DBI Lagrangian simplifies, 
\begin{equation}
{\cal L} = \sqrt{(1 - B_3^2) - \phi_t^2 +  \phi_x^2 } 
         = \sqrt{ 1 - B_3^2 } \sqrt{1 - \varphi_t^2 + \varphi_x^2},
\end{equation}
with 
\begin{equation}
       \varphi \equiv \frac{1}{\sqrt{1 - B_3^2}} \phi .
\end{equation}
  Consequently, for D-string, a constant $U(1)$ field strength tensor 
$F_{01}$, or, more generally, the gauge invariant combination, 
$B_3 \equiv B_{01} + F_{01}$, only contributes to a rescaling of the string 
tension and does not affect the dynamics.

  On the other hand, $B_1 = B_2 = 0$ and $B_3^2 = 1$ constitutes a singular 
case:
\begin{equation}
  {\cal L} = \sqrt{- \phi_t^2 +  \phi_x^2}.
\end{equation}
  This Lagrangian gives rise to the Bateman equation 
\begin{equation}
         \phi_t^2       \phi_{xx} 
     - 2 \phi_x  \phi_t \phi_{xt} 
     +   \phi_x^2       \phi_{tt} = 0,
\end{equation}
and the general solution is given by
\begin{equation}
F(\phi) t + G(\phi) x = $const$,
\end{equation}
where $F$ and $G$ are two arbitrary functions.
  See \cite{Baker:1999bf} for more discussion on this problem.
  
  Finally, let us emphasize the importance of a nonzero $B_3$ in this 
problem.
  If $B_3 = 0$, the DBI Lagrangian for D-string reduces to
\begin{equation}
{\cal L} = \sqrt{ 1
               - (1  +  B_1^2) \phi_t^2
               -  2 B_1 B_2    \phi_t  \phi_x
               + (1  -  B_2^2) \phi_x^2 }.
\end{equation}
  By a change of variables and rescaling,
\begin{equation}
  \left( \begin{array}{c}
          t' \\ x' 
         \end{array} \right) 
= \left( \begin{array}{rl} 
          a  & b \\ c & d
         \end{array} \right)
  \left( \begin{array}{c}
          t \\ x     
         \end{array} \right),
\end{equation}
\begin{equation}
  \tau   \equiv \frac{1}{\sqrt \lambda_-} t', \hspace{1cm}
  \sigma \equiv \frac{1}{\sqrt \lambda_+} x', 
\end{equation}
where $a,b,c,d$ and $\lambda_-, \lambda_+$ are solutions to the 
eigenvalue equations,
\begin{equation}
 \left( \begin{array}{cc}
    -( 1 + B_1^2) & - B_1 B_2 \\
    -  B_1 B_2  & 1 - B_2^2
 \end{array} \right) 
 \left( \begin{array}{cc}
    a & c \\
    b & d     
 \end{array} \right) =
 \left( \begin{array}{cc}
    a & c \\
    b & d     
 \end{array} \right)  
 \left( \begin{array}{cc}
   - \lambda_- & 0 \\
   0 &  \lambda_+
 \end{array} \right),
\end{equation}
we can transform the DBI Lagrangian with $B_3 = 0$ into the Nambu-Goto form, 
\begin{equation}
  {\cal L} = \sqrt{ 1 - \varphi_{\tau}^2 + \varphi_{\sigma}^2 }.
\end{equation}
  Consequently, the solutions of DBI equation with $B_3 = 0$ can be 
generated from solutions of the Nambu-Goto equation,
\begin{equation}
  (1 - \varphi_t^2) \varphi_{xx} + 2 \varphi_x \varphi_t \varphi_{xt}
- (1 + \varphi_x^2) \varphi_{tt} = 0,
\end{equation}
with 
\begin{equation}
  \phi(x,t) = \varphi(\tau(x,t), \sigma(x,t)).
\end{equation}

\subsection{Traveling wave solutions of the DBI equation}

  Since we shall be concerned with wave solutions, it is of interest to check 
if the DBI equation in $1+1$ dimension supports traveling waves. 
  For this purpose, we take a trial solution $\phi(x,t) = f (x - v \ t)$ 
and substitute it into the DBI equation Eq.(\ref{eq:dbi}).
  After some algebra, we get a quadratic equation for $v$, 
\begin{equation}
 (1 + B_1^2 - B_3^2) \ v^2 - 2 B_1 B_2 \ v - (1 - B_2^2 - B_3^2) = 0,
\end{equation}
with two roots,
\begin{eqnarray} \left\{
      \begin{array}{c} \alpha \\ \beta \end{array} \right\} &\equiv&
      \frac{      B_1     B_2 \pm \sqrt{B_1^2 B_2^2
            +(1 + B_1^2 - B_3^2)(1 - B_2^2 - B_3^2)}}
             {1 + B_1^2 - B_3^2}, \nonumber \\
       &=&\frac{  B_1     B_2 \pm \sqrt{
             (1 - B_3^2)(1 + B_1^2 - B_2^2 - B_3^2)}}
             {1 + B_1^2 - B_3^2}.
\label{eq:vel} 
\end{eqnarray}

  Several special cases are of interest for later discussions:
\begin{enumerate}
  \item $B_1 = 0$
   \begin{equation}
    \left\{
      \begin{array}{c} \alpha \\ \beta \end{array} \right\} \equiv
      \pm \sqrt \frac{1 - B_2^2 - B_3^2}{1 - B_3^2},
   \end{equation}

  \item $B_2 = 0$
   \begin{equation}
     \left\{
      \begin{array}{c} \alpha \\ \beta \end{array} \right\} \equiv
      \pm \sqrt \frac{1 - B_3^2}{1 + B_1^2 - B_3^2},
   \end{equation}

  \item $B_1 = B_2$
   \begin{equation}
    (i) \ \ |B_3|<1 \ \ \Rightarrow \
      \alpha = 1, \hspace{1cm}
      \beta  = \frac{-1 + B_1^2 + B_3^2}{\ \ 1 + B_1^2 - B_3^2},
   \end{equation}
   \begin{equation}
    (ii) \ \ |B_3|>1 \ \ \Rightarrow \
      \alpha = \frac{-1 + B_1^2 + B_3^2}{\ \ 1 + B_1^2 - B_3^2},   
      \hspace{1cm}
      \beta = 1.
   \end{equation}  
   Notice that only the first case $|B_3| <1$ satisfies causality constraint
   $|\alpha|,|\beta| \leq 1$.
   In view of this, we shall limit our discussion below with $|B_i| <1$.

   \item $B_1 = - B_2$ 

    This case amounts to a parity transformation of the above solutions, 
    $\alpha \rightarrow - \beta, \beta \rightarrow - \alpha$. 

\end{enumerate}

  The constant velocities $\alpha, \beta$ imply a linear dispersion 
relation $w(k) \propto \mbox{const.} k$ for plane wave solution 
$\phi (t,x;k) = {\cal A}(k) e^{i(k x - w(k) t)}$ and the fact that the wave 
frequency $w(k)$ is independent of the wave amplitude ${\cal A}(k)$, which 
presumably are consequences of general covariance of the DBI equation, 
distinguish this theory from other integrable systems, e.g., KdV equation 
and nonlinear Schroedinger equation \cite{Whitham:1974}.

\section{Scattering Wave Solutions to the Dirac-Born-Infeld Equation with
         Constant Background Fields}
  
  In this section, I shall follow the approach by \cite{Whitham:1974} in 
obtaining the wave solutions of DBI equation.

\subsection{Introduce wave-front coordinates and change of variables}
    
  In the usual solution of the linear wave equation, 
\begin{equation}
 \phi_{tt} - c^2 \phi_{xx} = 0,
\end{equation}
we define light-cone coordinates 
\begin{equation}
 \xi \equiv x - c \ t, \hspace{1cm} \eta \equiv x + c \ t,
\end{equation}
and the wave equation, in terms of the light-cone variables, is 
\begin{equation}
 \phi_{\xi\eta}=0.
\end{equation} 
  This transformed equation can be integrated and the solutions 
\begin{equation}
 \phi(x,t) = f(\xi) + g(\eta) = f(x - c \ t) + g(x + c \ t)
\end{equation} 
correspond to two non-interacting traveling waves with constant 
velocities $c, -c$. 

  We can generalize this method by defining the ``wave-front" coordinates, 
\begin{equation}
 \xi \equiv x - \alpha \ t, \hspace{1cm} \eta \equiv x - \beta \ t,
\label{eq:wfv}
\end{equation}
together with the inverse relations 
\begin{equation}
  x = \frac{\alpha \eta - \beta \xi}{\alpha-\beta}, \hspace{1cm} 
  t = \frac{\eta - \xi}{\alpha-\beta},
\end{equation}
where $\alpha$ and $\beta$ are traveling wave velocities of the DBI
equations, Eq.(\ref{eq:vel}). 

  For later convenience, it is useful to reduce the partial differential 
equation Eq.(\ref{eq:dbi}) into a first order form, and we introduce
\begin{equation}
 u \equiv \phi_\xi \equiv \frac{\partial \phi}{\partial \xi} 
   = u(\xi, \eta),
 \hspace{0.5cm}
 v \equiv \phi_\eta \equiv \frac{\partial \phi}{\partial \eta}
   = v(\xi, \eta). 
\end{equation}

  With these definitions, the original DBI equation Eq.(\ref{eq:dbi}) is 
then transformed into a coupled system of non-linear first order P.D.E.s,
\begin{eqnarray}
& & u_\eta = v_\xi, \hspace{1cm} \mbox{(integrable condition)} \nonumber \\
& &           (   p v - v^2 ) u_\xi
 + ( a + 2  u v - p u + q v ) u_\eta
            - (   q u + u^2 ) v_\eta = 0,
\label{eq:ndbi}
\end{eqnarray}
with
\begin{equation} 
        p \equiv  \frac{ 2 B_3 (B_1 - \alpha B_2) }
                         { ( \alpha - \beta ) (1 + B_1^2 - B_2^2) }, 
        \hspace{0.3cm}  
        q \equiv  \frac{ 2 B_3 (B_1 - \beta B_2)  }
                         { ( \alpha - \beta ) (1 + B_1^2 - B_2^2) },
        \hspace{0.3cm}
        a \equiv  \frac{ (1 + B_1^2 - B_3^2)}{(1 + B_1^2 - B_2^2)}.
\end{equation}

\subsection{Use of Hodograph transformation}

  Due to the general covariant property of the DBI Lagrangian, one can not 
transform the nonlinearity away by performing a coordinate transformation. 
  Therefore, we introduce the Hodograph transformation \cite{Whitham:1974} to
attack this problem.
  The general idea of the Hodograph transformation is to exchange the roles
of independent variables ($\xi, \eta$) and dependent variables ($u,v$) in 
the P.D.E. such that it is possible to obtain linearized equations. 
  In the case of two independent variables, through the use of chain 
rules, we have
\begin{equation}
 {\large
\begin{array}{cc}
     1 = \frac{\partial u}{\partial u} |_v  
       = \frac{\partial u}{\partial \xi}  \frac{\partial \xi}{\partial u}
       + \frac{\partial u}{\partial \eta} \frac{\partial \eta}{\partial u},
       & \hspace{0.9cm}
     0 = \frac{\partial u}{\partial v} |_u
       = \frac{\partial u}{\partial \xi}  \frac{\partial \xi}{\partial v}
       + \frac{\partial u}{\partial \eta} \frac{\partial \eta}{\partial v},
       \\
     0 = \frac{\partial v}{\partial u} |_v
       = \frac{\partial v}{\partial \xi}  \frac{\partial \xi}{\partial u}
       + \frac{\partial v}{\partial \eta} \frac{\partial \eta}{\partial u},
       & \hspace{0.9cm}
     1 = \frac{\partial v}{\partial v} |_u
       = \frac{\partial v}{\partial \xi}  \frac{\partial \xi}{\partial v}
       + \frac{\partial v}{\partial \eta} \frac{\partial \eta}{\partial v} .
   \end{array} }
   \end{equation}
Or, in a matrix notation,
\begin{equation}
  \left( {\large
  \begin{array}{cc}  
   \frac{\partial u}{\partial \xi} & \frac{\partial u}{\partial \eta}\\
   \frac{\partial v}{\partial \xi} & \frac{\partial v}{\partial \eta}
  \end{array} }
  \right)
  \left( {\large
  \begin{array}{cc}
   \frac{\partial \xi}{\partial u}  & \frac{\partial \xi}{\partial v}\\
   \frac{\partial \eta}{\partial u} & \frac{\partial \eta}{\partial v}
  \end{array} }
  \right) =   
  \left(
    \begin{array}{cc}
     1 & 0 \\
     0 & 1
    \end{array}
  \right).
\end{equation}
  One can reverse the matrix equation and replace the field variables
$u_\xi , u_\eta, v_\xi$, and $v_\eta$ in terms of 
$\xi_u , \xi_v , \eta_u , \eta_v$,
\begin{equation} 
    \left( {\large
    \begin{array}{cc}
     \frac{\partial \xi}{\partial u}  & \frac{\partial \xi}{\partial v} \\
     \frac{\partial \eta}{\partial u} & \frac{\partial \eta}{\partial v}
    \end{array} }
    \right)  = \frac{1}{\Delta}
    \left(  {\large
    \begin{array}{rr}
     \frac{\partial v}{\partial \eta} & - \frac{\partial u}{\partial \eta}\\
   - \frac{\partial v}{\partial \xi}  &   \frac{\partial u}{\partial \xi}
    \end{array} }
    \right), \hspace{1cm}   
    \Delta \equiv \det
    \left| {\large \begin{array}{rr}
    \frac{\partial u}{\partial \xi} & \frac{\partial u}{\partial \eta} \\
    \frac{\partial v}{\partial \xi} & \frac{\partial v}{\partial \eta}
    \end{array} } \right| .
   \end{equation}

   After applying the Hodograph transformation to Eq.(\ref{eq:ndbi}), we 
obtain a linear system of coupled P.D.E.s,
\begin{equation}
  \begin{array}{l}
    \xi_v = \eta_u \hspace{1cm}
      \mbox{ (integrable condition)}\\
                          (   q u +   u^2 ) \xi_u
              + ( a + 2 u v - p u + q v   ) \xi_v
                         - (  p v -   v^2 ) \eta_v = 0
   \end{array}
\label{eq:dbi2}  
\end{equation}
or, equivalently, two linear second order P.D.E.s,
\begin{eqnarray}
                       ( q u + u^2) \xi_{uu}
         + ( a + 2 u v - p u + q v) \xi_{uv}
         +           ( - p v + v^2) \xi_{vv} 
         +               ( q + 2 u) \xi_u  
         +             ( - p + 2 v) \xi_v = 0, \\
                       ( q u + u^2) \eta_{uu}
         + ( a + 2 u v - p u + q v) \eta_{uv}
         +            (- p v + v^2) \eta_{vv}
         +               ( q + 2 u) \eta_u
         +             ( - p + 2 v) \eta_v = 0.
\end{eqnarray}

  These equations are then subject to the standard treatment for a 
solution of hyperbolic differential equations. 
 
\subsection{Solutions of characteristic equations} 

  To solve the linear P.D.E. Eq.(\ref{eq:dbi2}), we first need to identify 
the characteristics of the hyperbolic systems and change to 
characteristic variables in order to obtain decoupled wave equations.
  The equations which characteristic curves satisfy are extracted from 
the quadratic part of the linear P.D.E.. Namely,
\begin{equation}
                  (q u + u^2)    dv^2 
   - ( a + 2 u v - p u + q v) dv du 
               + ( v^2 - p v)    du^2 = 0, 
\end{equation}
\begin{equation}
   \Rightarrow 
   \left( \frac{du}{dv} \right)_{\pm} =
          \frac{(a + 2 u v - p u + q v) \pm
          \sqrt{(a + 2 u v - p u + q v)^2 - 4 (u^2 + q u)(v^2 - p v)} }
               {2(v^2 - p v)}. 
\label{eq:cha}
\end{equation}

  These characteristic curves are exactly integrable and the solutions are 
given by 
\begin{eqnarray}
  2 (u - r^2 v) =     - (q + p r^2) + 
                  \sqrt{(q + p r^2)^2 + 4 a r^2}, \\
  2 (v - s^2 u) =   \ \ (p + q s^2) +
                  \sqrt{(p + q s^2)^2 + 4 a s^2},
\end{eqnarray}
where $r,s$ are characteristic variables, which come as integration 
constants along the characteristic curves Eq.(\ref{eq:cha}). 

  The defining equations given above allow us to generate relations 
between various derivatives, e.g.,  
\begin{eqnarray}
   \frac{\partial r}{\partial u} =   \frac{1}{r} h(r,s) f_1 (r), 
   \hspace{0.3cm} 
   \frac{\partial r}{\partial v} =           - r h(r,s) f_1 (r), 
   \hspace{0.3cm}
   \frac{\partial s}{\partial u} =           - s h(r,s) f_2 (s), 
   \hspace{0.3cm}
   \frac{\partial s}{\partial v} =   \frac{1}{s} h(r,s) f_2 (s),
\end{eqnarray}
where
\begin{eqnarray}
 f_1 (r) &\equiv& \sqrt{(q + p r^2)^2 + 4 a r^2}, \hspace{2cm}
 f_2 (s)  \equiv  \sqrt{(p + q s^2)^2 + 4 a s^2}, \nonumber \\
 g(r,s)  &\equiv& p^2 r^2 + q^2 s^2 + (p q + 2 a )(1 + r^2 s^2),  
 \hspace{1cm}
 h(r,s)  \equiv  \frac{1 - r^2 s^2}{g + f_1 f_2}.
\end{eqnarray}

  Rewrite the first order P.D.E. Eq.(\ref{eq:dbi2}) in terms of the 
characteristic variables  
$\xi(u,v) \rightarrow \xi(r,s), \eta(u,v) \rightarrow \eta(r,s)$, we have
\begin{equation}
              \eta_r + r^2 \xi_r = 0, \hspace{1cm}  
          s^2 \eta_s +     \xi_s = 0.
\end{equation}
  These equations imply two decoupled wave equations,
\begin{equation}
      \xi_{rs} = \eta_{rs} = 0, 
\end{equation} 
and the solutions are
\begin{eqnarray}
 \xi(r,s) &=& F(r) - \int_{s_0}^s {\tilde s}^2 G'(\tilde s) d \tilde s,
 \nonumber \\
 \eta(r,s) &=& G(s)+ \int^{r_0}_r {\tilde r}^2 F'(\tilde r) d \tilde r,
\label{eq:cha3}
\end{eqnarray}
where $F(r)$ and $G(s)$ are arbitrary functions and they are only 
well-defined up to constants, which are specified by the choices of 
$s_0$ and $r_0$.

  Given these decoupled wave solutions, we can derive the original wave 
solution in terms of the characteristic variables, 
$\phi(x,t) \ \Rightarrow \ \phi(\xi, \eta) \ 
\Rightarrow \phi(u,v) \ \Rightarrow \ \phi(r,s)$, 
 \begin{eqnarray}
        \phi_r = \phi_\xi \xi_r + \phi_\eta \eta_r
              &=&  ( u - r^2 v ) F'(r) 
               =  \frac{1}{2} 
                  \left[ f_1 - \sqrt{f_1^2 - 4 a r^2} \right] F'(r), \\
        \phi_s = \phi_\xi \xi_s + \phi_\eta \eta_s
              &=&  ( v - s^2 u ) G'(s)   
               =  \frac{1}{2} 
                  \left[ f_2 + \sqrt{f_2^2 - 4 a s^2} \right] G'(s), \\
\phi = \phi(r,s) &=& \int \left[ \sqrt{(q + p r^2)^2 + 4 a r^2} 
                                     - (q + p r^2) \right] F'(r) dr 
       \nonumber \\
                 &+&  \int \left[ \sqrt{(p + q s^2)^2 + 4 a s^2}
                                      + (p + q s^2) \right] G'(s) ds . 
\label{eq:ans} 
\end{eqnarray}

 \subsection{Scattering wave solutions in the space-time representation}

  After various transformations, we arrive at a solution in terms of the 
characteristic variables $\phi = \phi(r,s)$, Eq.(\ref{eq:ans}).
  To study the time evolution and identify physical observables, we need to
re-express the wave solution in terms of space-time variables.
  First of all, we define the shape variables,
 \begin{equation}
          \rho \equiv F(r) \Rightarrow r = F^{-1} (\rho), \hspace{1cm}
        \sigma \equiv G(s) \Rightarrow s = G^{-1} (\sigma),
 \label{eq:sha}
 \end{equation} 
and replace the functions $F(r), G(s)$ by
\begin{eqnarray}
\Phi_1 (\rho)   &\equiv& \frac{1}{2} \int
\left[ f_1 - \sqrt{f_1^2 - 4 a r^2} \right] d \rho,  
\label{eq:sha2} \\
\Phi_2 (\sigma) &\equiv& \frac{1}{2} \int
\left[ f_2 + \sqrt{f_2^2 - 4 a s^2} \right] d \sigma. 
\label{eq:sha3}
\end{eqnarray}
  In terms of the shape variables $\rho, \sigma$ and the new functions 
$\Phi_1,\Phi_2$, the wave solution can be written in a decoupled form, 
\begin{equation}
  \phi = \Phi_1 (\rho) + \Phi_2 (\sigma).
\end{equation}
  
  We can now relate the space-time variables $(x,t)$ to the shape 
variables $(\rho, \sigma)$, using Eqs.(\ref{eq:wfv}) (\ref{eq:cha3})
(\ref{eq:sha}),
\begin{eqnarray}
          & & x - \alpha \ t = \xi = \rho -
             \int_{-\infty}^{\sigma}
             \frac{{\Phi'}_2^2 - p \Phi'_2}{a + q \Phi'_2}
             \ d \tilde \sigma , 
\label{eq:sha4}\\
          & & x - \beta \ t = \eta = \sigma +
             \int_{\rho}^{\infty}
             \frac{{\Phi'}_1^2 + q \Phi'_1}{a - p \Phi'_1} 
             \ d \tilde \rho ,
\label{eq:sha5}
\end{eqnarray}
where we have solved $r^2, s^2$ in terms of $\Phi'_1$ and $\Phi'_2$
through Eqs.(\ref{eq:sha2}) (\ref{eq:sha3}).
  Finally, we derive time and space as functions of the shape variables
\begin{eqnarray}
          t &=& \frac{1}{\alpha - \beta} \left[ \ \ \sigma - \rho
              \ \ + \int_{\rho}^{\infty} \
                \frac{{\Phi'}_1^2 + q \Phi'_1}{a - p \Phi'_1} 
                \ d \tilde \rho
              \ \ + \int_{-\infty}^{\sigma} \
                \frac{{\Phi'}_2^2 - p \Phi'_2}{a + q \Phi'_2} 
                \ d \tilde \sigma \right], 
\label{eq:sha6} \\
          x &=& \frac{1}{\alpha - \beta}
                    \left[ \alpha \sigma - \beta \rho
              + \alpha \int_{\rho}^{\infty}
                \frac{{\Phi'}_1^2 + q \Phi'_1}{a - p \Phi'_1} 
                \ d \tilde \rho
              + \beta \int_{-\infty}^{\sigma}
                \frac{{\Phi'}_2^2 - p \Phi'_2}{a + q \Phi'_2} 
                \ d \tilde \sigma \right].
\label{eq:sha7}
\end{eqnarray}
  Solving the inverse relations to determine the functional form of
shape variables in terms of space-time variables $\rho = \rho(x,t), 
\sigma = \sigma(x,t)$, we then have a complete evolution of the 
scattering event, represented by $\phi(x,t)$ through Eq.(\ref{eq:sha4}) 
(\ref{eq:sha5}) (\ref{eq:sha6}) (\ref{eq:sha7}),
\begin{equation}
\phi(x,t) = \Phi_1 (x - \alpha \ t + \int_{-\infty}^{\sigma (x,t)}
             \frac{{\Phi'}_2^2 - p \Phi'_2}{a + q \Phi'_2}
             \ d \tilde \sigma) \
          + \Phi_2 (x - \beta \ t - \int_{\rho (x,t)}^{\infty}
             \frac{{\Phi'}_1^2 + q \Phi'_1}{a - p \Phi'_1}
             \ d \tilde \rho).
\label{eq:wave}
\end{equation}

\subsection{D-string profiles as scattering events}
  
  The space-time representation of wave solutions to the DBI equation, 
Eq.(\ref{eq:wave}), strongly suggests an interpretation for the time 
evolution of D-string motion as a scattering event.
  For instance, if we identify two localized functions $\Phi_1, \Phi_2$ as 
interacting wave packets, and assume that they are nonzero only within  
finite intervals,
\begin{equation}
 \Phi_1(\rho)   \not = 0, \hspace{0.5cm} -2 \leq \rho   \leq 0; \hspace{1cm} 
 \Phi_2(\sigma) \not = 0, \hspace{0.5cm}  0 \leq \sigma \leq 2, 
\label{eq:bump}
\end{equation}
then one can interpret the wave solution Eq.(\ref{eq:wave}) at $t \rightarrow
+ \infty$ as a scattered final state with phase shifts 
\begin{equation}
\chi_1 \equiv \int_{-\infty}^{\infty}
 \frac{{\Phi'}_2^2 - p \Phi'_2}{\ a + q \Phi'_2} 
\ d \tilde \sigma, \hspace{1cm}
\chi_2 \equiv \int_{-\infty}^{\infty}
 \frac{{\Phi'}_1^2 + q \Phi'_1}{\ a - p \Phi'_1}
\ d \tilde \rho,
\label{eq:chi}
\end{equation} 
which originates from two incident waves with constant velocities
$\Phi_1 (x - \alpha \ t)$ and $\Phi_2 (x - \beta \ t)$ at 
$t \rightarrow - \infty$.

  To see this, examing the long time behavior of Eqs.(\ref{eq:sha4}) 
(\ref{eq:sha5}), we find that as long as localized wave packets are 
well-behaved (meaning the integrals in Eqs.(\ref{eq:sha4}) (\ref{eq:sha5}) are
finite for any value of $\rho$ and $\sigma$), both $\rho$ and $\sigma$ have to
diverge, 
\begin{equation}
    t   \rightarrow + \infty   \ \  \Rightarrow \ \ 
 \rho   \rightarrow - \infty , \hspace{1cm}
 \sigma \rightarrow + \infty   \hspace{1cm} (\alpha > 0, \beta < 0).
\end{equation}
  Therefore, the transformation rules between $\rho, \sigma$ and $x,t$
simplifies in the $t \rightarrow + \infty$ limit, and the final state
wave solution converges to
\begin{equation}
 \phi(x,t \rightarrow + \infty) = \Phi_1 (x - \alpha \ t + \chi_1) 
                                + \Phi_2 (x - \beta  \ t - \chi_2).
\end{equation}
  
  On the other hand, we see from Eq.(\ref{eq:sha6}) that $\sigma$ must be 
less than $\rho$ for sufficient large negative time $t = - T_0$, and before 
that moment there is no overlap between two wave packets $\Phi_1, \Phi_2$ 
due to constraint Eq.(\ref{eq:bump}).
  Therefore, we can write the initial state wave solution as a sum of
two independent incident wave packets,
\begin{equation}
 \phi(x,t \rightarrow - \infty ) = \Phi_1 (x - \alpha \ t)
                                 + \Phi_2 (x - \beta  \ t).
\end{equation}
  
  To summarize, our general solutions show that the scattering event only 
induces phase shifts on incoming wave packets and both the wave velocities and
the wave profiles are retained after the scattering.
  While there is no qualitative difference (except for the singular cases 
which are ignored in this study) between the DBI theory with nonzero 
constant background fields and the zero field case for the string dynamics, 
our solutions give explicit dependences of wave velocities and phase shifts on 
the external field parameters. 
  Finally, let us comment that it is possible to verify these descriptive
pictures by choosing a particular form of $\Phi_1, \Phi_2$ and solve for the
whole scattering event.
 
\subsection{Scattering phase shifts and the energy momentum of D-string}

  It is interesting to notice that the phase shift of each scattering 
wave is entirely induced by the wave profile of the incident company,
Eq.(\ref{eq:chi}). 
  As the scattering solutions demonstrate certain ``solitonic" behaviors,
we are curious to know if one can apply any physical interpretation to these 
phase shifts.
  Indeed, Barbashov and Chernikov have pointed out that the phase shifts of 
scattering waves can be related to the energies (or momenta) of the incoming 
companies in the case of pure Nambu-Goto theory.
  Given the analytic expressions of the phase shifts in terms of the 
incoming wave profiles and constant background fields, we wish to examine if
these connections persist with nonzero backgrounds.
 
  However, a complete solution is beyond our ability and I can only illustrate
the physical meaning of the phase shifts $\chi_1, \chi_2$ in several special 
cases. 
  For this purpose, we change the definition of the DBI Lagrangian to 
\begin{equation}
 {\cal L}_0 \equiv \sqrt{1 - B_3^2} - {\cal L},
\end{equation}
  and the canonical momentum is given by
\begin{equation} 
  \Pi \equiv \frac{\delta {\cal L}_0}{\delta \phi_t} 
         = - \frac{\delta {\cal L}}  {\delta \phi_t} 
         =   \frac{1}{{\cal L}} \
           [ \ (1 + B_1^2) \phi_t + B_1 B_2 \phi_x - B_1 B_3 \ ]. 
\end{equation} 
   The conserved quantities, energy and momentum can be deduced, 
\begin{eqnarray}
 {\cal H} &\equiv& \phi_t \Pi - {\cal L}_0 \nonumber \\
 &=& \ \frac{1}{{\cal L}} \ [ \ (1 - B_2^2) \phi_x^2 
                         - B_1 B_2 \phi_x \phi_t + 
     2 B_2 B_3 \phi_x + B_1 B_3 \phi_t + (1 - B_3^2) \ ] - 
     \sqrt{1 - B_3^2}, 
\label{eq:enr}\\
   {\cal P} &\equiv& \phi_x \Pi  
   = \ \frac{1}{{\cal L}} \  
   [ \ (1 + B_1^2) \phi_x \phi_t + B_1 B_2 \phi_x^2 - B_1 B_3 \phi_x \ ].
\label{eq:mom}
\end{eqnarray}
 
   I compare the formulae of the phase shifts $\chi_1, \chi_2$, 
Eq.(\ref{eq:chi}), with the momentum and the Hamiltonian density by 
substituting the traveling wave solutions 
\[ \Phi_1 (x,t) = f(x - \alpha \ t), \hspace{1cm}
   \Phi_2 (x,t) = g(x - \beta  \ t) \] 
into Eqs.(\ref{eq:enr}) (\ref{eq:mom}) with
\[ {\cal H}_i \equiv {\cal H} (\Phi_i), \hspace{1cm}
   {\cal P}_i \equiv {\cal P} (\Phi_i). \]
The results for the following special cases are: 
\begin{enumerate}
 \item $B_1=0$
   \begin{equation} 
   \chi_2 = \int \frac{1}{\sqrt{1 - B_3^2}} {\cal H}_1, \hspace{1cm}
   \chi_1 = \int \frac{1}{\sqrt{1 - B_3^2}} {\cal H}_2. 
   \end{equation}

 \item $B_2=0$
   \begin{equation}
   \chi_2 = \int \frac{1}{\sqrt{1 +  B_1^2 - B_3^2}} {\cal P}_1, 
\hspace{1cm}
   \chi_1 = \int \frac{1}{\sqrt{1 +  B_1^2 - B_3^2}} {\cal P}_2.
   \end{equation}

 \item $B_1 = B_2$ 
   \begin{equation}
   \chi_2 = \int \frac{\sqrt{1 - B_3^2}}{2 (1 +  B_1^2 - B_3^2)}
            ({\cal H}_1 - {\cal P}_1), \hspace{1cm}
   \chi_1 = \int \frac{1}{\sqrt{1 - B_3^2}} 
            ({\cal H}_2 + {\cal P}_2).
   \end{equation}
\end{enumerate}

\section{Summary and Conclusion}

  In this work, I discuss the scattering wave solutions of the 
Dirac-Born-Infeld equation in the presence of constant background fields.
  Using the Hodograph transformation and various change of variables, we 
are able to reduce the DBI equation into a system of linear partial 
differential equations and obtain exact wave solutions propagating along
a D-string.
  Under suitable conditions, the wave solutions we derived can be interpreted
as scattering events for localized wave packets.  
  Our solutions show that the effect of constant background fields only change
the traveling velocities of the incoming waves from the speed of light and 
the final result of the scattering is to induce phase shifts on outgoing waves
without modifying their wave profiles.
  The phase shifts of scattering waves can be related to the energies and 
momenta of individual wave packets in the presence of constant background 
fields.
  Finally, we give explicit relations between phase shifts and 
energy-momentum of D-string for several special cases of background field 
parameters.
  
  This work is supported by NSC of Taiwan for the research project of 
``Investigations of some non-perturbative phenomena in physics" under the 
budget number NSC 88-2112-M-002-001-y. I was benefited from many discussions 
with Pei-Ming Ho, Miao Li and Yeong-Chuan Kao.

\end{document}